\documentclass[envcountsame]{llncs}

\usepackage[obeyFinal]{todonotes}
\usepackage{xspace}
\usepackage{amsfonts,amssymb,amsmath}
\usepackage[scr]{rsfso}
\usepackage[ruled, vlined, linesnumbered]{algorithm2e}
\usepackage{cite}

\newcommand{\Amc}{\ensuremath{\mathcal{A}}\xspace}
\newcommand{\Cmc}{\ensuremath{\mathcal{C}}\xspace}
\newcommand{\Gmc}{\ensuremath{\mathcal{G}}\xspace}
\newcommand{\Mmc}{\ensuremath{\mathcal{M}}\xspace}
\newcommand{\Nmc}{\ensuremath{\mathcal{N}}\xspace}
\newcommand{\Omc}{\ensuremath{\mathcal{O}}\xspace}

\newcommand{\Cmf}{\ensuremath{\mathfrak{C}}\xspace}
\newcommand{\Omf}{\ensuremath{\mathfrak{O}}\xspace}

\newcommand{\powfin}{\ensuremath{\mathscr{P}_\mathsf{fin}}\xspace}

\newcommand{\SROIQ}{\ensuremath{\mathcal{SROIQ}}\xspace}
\newcommand{\EL}{\ensuremath{\mathcal{E\!L}}\xspace}
\newcommand{\ALC}{\ensuremath{\mathcal{ALC}}\xspace}

\newcommand{\isjust}{\textsc{is-just}\xspace}
\newcommand{\alljust}{\textsc{all-just}\xspace}

\newcommand{\NP}{\textsc{NP}\xspace}
\newcommand{\PT}{\textsc{P}\xspace}
\newcommand{\PSpace}{\textsc{PSpace}\xspace}

\title{Axiom Pinpointing}
\author{Rafael Pe\~naloza}
\institute{IKR3 Research Lab, University of Milano-Bicocca, \\ \url{rafael.penaloza@unimib.it}}

\begin{document}

\maketitle

\begin{abstract}
Axiom pinpointing refers to the task of finding the specific axioms in an ontology which are responsible for a consequence
to follow. This task has been studied, under different names, in many research areas, leading to a reformulation and
reinvention of techniques. In this work, we present a general overview to axiom pinpointing, providing the basic 
notions, different approaches for solving it, and some variations and applications which have been considered in 
the literature. This should serve as a starting point for researchers interested in related problems, with an ample bibliography
for delving deeper into the details.
\end{abstract}

\section{Introduction}

Intelligent applications need to represent and handle knowledge effectively. For that reason, many different knowledge
representation languages
have been developed, providing formal semantics and reasoning methods for
deriving implicit consequences from explicitly represented elements (also called axioms). As these knowledge bases or ontologies 
grow, they become harder to maintain and verify and when---inevitably---errors occur, they are harder to understand and correct. 
Indeed, it is nowadays common to encounter knowledge bases with tens-of-thousands of axioms, and detecting the handful 
responsible for a given consequence would be impossible without the help of an automated tool.

Axiom pinpointing refers to the task of identifying the axioms in a knowledge base that are responsible for a given 
consequence. Assuming that the representation language is monotonic (that is, adding new knowledge does not remove
any previous consequences), a relevant set of axioms is nothing more than a subset-minimal subontology which still entails
the consequence under consideration. Such a set is called a justification. It is not difficult to see (and will be exemplified in
the following sections) that there may exist multiple justifications for a single consequence. It is thus important to try to compute
them all, for a full understanding of the derivation.

Understanding the causes of a consequence are not just instrumental for knowledge engineers to understand the ontologies
they work on. Axiom pinpointing is also a relevant step for repairing modeling errors which could have been understood
through misunderstanding, automated knowledge extraction, or merely typing or methodological errors. Identifying the
potentially faulty axioms is the first step towards correcting the error. A third usecase for axiom pinpointing is explainability:
if an intelligent system makes a decision based on a reasoning process, it is important to be able to explain the reasoning
behind this decision to all the stakeholders involved.

We provide a general overview of axiom pinpointing over many different representation languages. Although
we use terminology and results primarily developed in the context of description logics \cite{dlhandbook}, we try to keep
the presentation as general as possible to include other well-known monotonic formalisms like databases, logic programming,
and propositional logic. Our goal is to describe what the main reasoning tasks associated to axiom pinpointing are, provide
the basic templates for solving them, and present a few variants and applications from the literature. The hope is that this
general description serves as a first step towards a unified description of the tasks for different areas of knowledge representation,
and aids in a common development of new methods and tools.

\section{Axiom Pinpointing}
\label{sec:ap}

To make the presentation as general as possible, we consider an abstract notion of an \emph{ontology language}, which has 
four components: a class \Amc of well-formed \emph{axioms}; a class \Cmc of \emph{consequences}; a class 
$\Omf\subseteq\powfin(\Amc)$ of valid \emph{ontologies}, where $\powfin(X)$ is the class of all \emph{finite} subsets of $X$, 
such that if $\Omc\subseteq \Omc'$ and $\Omc'\in\Omf$, then $\Omc\in\Omf$ (that is, every subset of an ontology is also an ontology); 
and an \emph{entailment relation} ${\models}\subseteq \Omf\times\Cmc$, expressed in infix notation, such that for every 
two ontologies 
$\Omc,\Omc'\in\Omf$ and consequence $c\in\Cmc$, if $\Omc\models c$ and $\Omc\subseteq\Omc'$, then $\Omc'\models c$;
that is, the entailment relation must be monotonic w.r.t.\ the ontology.

We note that in some existing work on axiom pinpointing and related topics, an ontology is often defined to be just a finite set of axioms, 
which is a special case of our definition.
We decided to use this more general notion to account for syntactic restrictions that are common in description logics. For example,
it allows for acyclic TBoxes, but also for the syntactic restrictions imposed to guarantee decidability of reasoning in \SROIQ 
ontologies~\cite{HKS06-sroiq,dlhandbook}.
It also allows for other languages not typically considered ontological, such as propositional formulas in conjunctive normal form
(CNF) or constraint satisfaction problems, to name just two examples.

\begin{figure}[tb]
\centering
\includegraphics[width=0.95\textwidth]{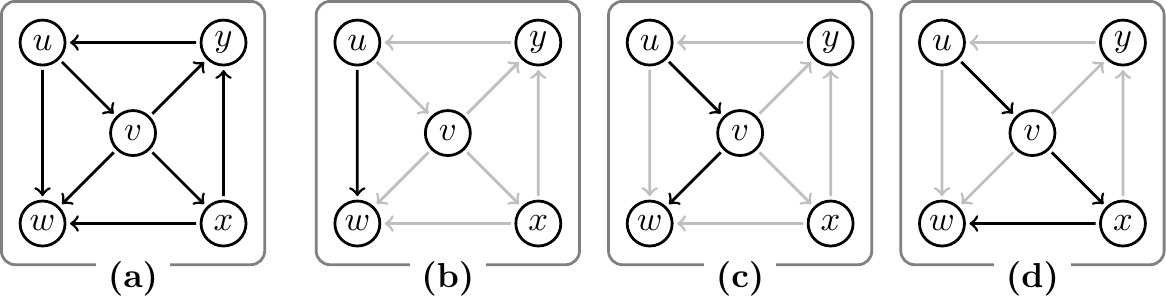}
\caption{The ontology \Gmc depicted as a graph {\bf (a)}, and three justifications for the consequence $(u,w)$ {\bf (b)--(d)}.}
\label{fig:exagraph}
\end{figure}

To aid the understanding of the notions presented here, we will use a very simple ontology language dealing with reachability
in finite graphs. Specifically, given a countable set $V$ of \emph{vertices}, let $\Amc=\Cmc=\{(v,w)\mid v,w\in V\}$; that is, axioms and
consequences are given by ordered pairs of vertices (called \emph{edges}), and $\Omf=\powfin(\Amc)$. Intuitively, an ontology is a 
finite graph, and $\Omc\models (v,w)$ iff $w$ is reachable from $v$ in the graph \Omc.
For example, the graph \Gmc in Figure~\ref{fig:exagraph}~(a) is one such ontology, and $\Gmc\models (u,w)$ but 
$\Gmc\not\models(w,u)$.
Importantly, this is only an \emph{example} of a very simple ontology language, but many of the intuitions obtained from it
apply also to more complex langauges.

For any given ontology language, the main reasoning task is to decide \emph{entailments}. Formally, given an ontology
$\Omc\in\Omf$ and a consequence $c\in\Cmc$, we are interested in deciding whether $\Omc\models c$ holds. Depending on the
specific language considered, different methods can be developed to solve this reasoning task, and its computational complexity 
may vary. In fact, already within the family of description logics, we can find examples where the complexity of entailment checking
varies from polynomial time up to doubly-exponential time. Moreover, there exist ontology languages with an undecidable entailment
problem, but for the scope of this work we focus only on cases where this problem is decidable with complexity \Cmf.
Once that adequate methods for deciding entailments have been developed, one is often interested in solving more complex
reasoning tasks.

Axiom pinpointing is a non-standard reasoning task, which focuses on identifying the axioms that are responsible for a consequence
to follow from an ontology. Formally, \emph{axiom pinpointing} is the task of identifying one or all the justifications for a given
consequence, where a justification is a minimal sub-ontology that still entails the consequence.
\begin{definition}[justification]
\label{def:just}
Let \Omc be an ontology and $c$ a consequence such that $\Omc\models c$. The sub-ontology $\Mmc\subseteq\Omc$ is called a
\emph{justification} for $c$ w.r.t.\ \Omc iff (i)~$\Mmc\models c$ and (ii) for every $\Mmc'\subsetneq\Mmc$, $\Mmc'\not\models c$.
\end{definition}
Here we are using the standard name from description logics, but it is worth noting that justifications are known with
different names by different communities. For example, they are also known as MUSes in SAT \cite{LiSa08}, MESCs in 
CSP \cite{MeMa14}, causes in DBs \cite{MGHKMS10}, and MIPS, MUPS, and MinAs in DLs \cite{ScCo-03,Pena:PhD}.

Consider for example the graph \Gmc from Figure~\ref{fig:exagraph}~(a), which entails the consequence $(u,w)$;
that is, $w$ is reachable from $u$. This consequence has three justifications, which are the three subgraphs (b)--(d) of the same
figure. Note that there exist other sub-ontologies that still entail $(u,w)$, but these contain at least one of the justifications depicted,
and hence are not subset-minimal.
In general, when an ontology corresponds to a graph, a justification for the consequence $(u,v)$ is a simple path from $u$ to $v$.

Technically, it is not difficult to come out with algorithms which compute one or all justifications. The simplest approach is to
exploit the existing reasoners and find justifications through repeated entailment checks. This approach is known as
\emph{black-box} because it uses the reasoner without any modification. A black-box method for computing one justification
is described in Algorithm~\ref{alg:one}.
\begin{algorithm}[tb]
\DontPrintSemicolon
\KwData{Ontology \Omc, consequence $c$}
\KwResult{A justification \Mmc for $c$ w.r.t.\ \Omc}
$\Mmc \gets \Omc$ \;
\For{$\alpha\in\Omc$}{
  \If{$\Mmc\setminus\{\alpha\}\models c$}{ \label{line:if}
    $\Mmc\gets \Mmc\setminus\{\alpha\}$
  }
}
{\bf Return} \Mmc
\caption{Compute one justification}
\label{alg:one}
\end{algorithm}
It tries to remove each axiom from the ontology, as long as it still entails the consequence. An invariant of the {\bf for} loop is that
$\Mmc\models c$, hence the resulting set satisfies the first condition from Definition~\ref{def:just}. Moreover, by monotonicity of 
$\models$ the resulting set cannot have any superfluous axioms, meaning that it is a justification. The resulting justification 
obtained through this algorithm depends on the order in which the axioms are selected for removal, but independently of the 
order, each axiom needs to be tested exactly once.

To find \emph{all} justifications, one can simply enumerate all possible sub-ontologies $\Mmc\subseteq\Omc$ and verify for each
of them that they entail $c$ and that no strict subset entails $c$ as well. Any such \Mmc is a justification. This method requires 
to check exponentially many times (i.e., once for each subset of \Omc) whether an ontology is a justification. Obviously, the
number of checks can be greatly reduced simply by taking into account the monotonicity of the entailment relation: if a set
$\Omc'\subseteq\Omc$ is such that $\Omc'\models c$, then there is no need to control any strict superset of $\Omc'$. Indeed, 
any such superset still entails $c$, but will not be minimal. Conversely, if $\Omc'\not\models c$, then no subset of $\Omc'$ can be
a justification because it cannot entail $c$. Other optimisations can be considered. For example, \cite{KPSH05} presents an
enumeration method based on Reiter's Hitting Set Tree method~\cite{Reit87}, which guides the search for new justifications. However, 
as we will see later, one cannot avoid testing exponentially many sets. In the following section we analyse this, and other complexity
issues in more detail.

\section{Complexity}

In this section, we consider computational complexity issues related to axiom pinpointing. We have already seen two basic
algorithms for computing one or all justifications, but it remains unclear whether more advanced techniques can provide 
improvements in terms of worst-case complexity. Although our goal is to understand the computational problem of \emph{finding}
the justifications, to analyse the complexity we consider their decision variants.

\begin{definition}[\isjust,\alljust]
\isjust is the problem of deciding, given an ontology \Mmc and a consequence $c$, whether \Mmc is a justification of $c$. 
\alljust is the problem of deciding, given an ontology \Omc, a consequence $c$, and a set $\{\Mmc_1,\ldots,\Mmc_n\}$ of 
sub-ontologies of \Omc, whether $\Mmc_1,\ldots,\Mmc_n$ are \emph{all} the justifications of $c$ w.r.t.\ \Omc.
\end{definition}

\subsection{One Justification}

Recall that the first condition for a justification is that it still entails the consequence $c$. Hence, entailment is a sub-problem
of \isjust. In particular this means that \isjust is necessarily at least as hard as entailment. More formally, if entailment is hard
for the complexity class \Cmf, then \isjust must be \Cmf-hard as well. Note also that Algorithm~\ref{alg:one} can be easily
modified to decide \isjust as well: if the test in line \ref{line:if} succeeds at any point, then the input ontology \Omc is \emph{not}
a justification and the method can exit with failure; if the whole loop runs without exiting, then \Omc was a justification.
In the worst case, this algorithm has to make one call to the entailment test for each axiom in \Omc. This means that \isjust
can be decided through polynomially many entailment tests; that is, \isjust is in $\PT^\Cmf$, if entailment is in \Cmf. Note that
if \Cmf is at least \PSpace, then the polynomial enumeration can be absorbed into the \Cmf oracle.

\begin{proposition}
\label{prop:one:same}
If entailment is \Cmf-complete for \Cmf at least \PSpace, then \isjust is \Cmf-complete as well.
\end{proposition}
The consequence of this proposition is that there is no need to analyse the complexity of \isjust for expressive ontology languages
with complex entailment relations. However, the black-box method for deciding this problem still leaves a gap when the complexity
of entailment \Cmf is below \PSpace; in those cases, \Cmf is usually smaller than $\PT^\Cmf$ and hence \Cmf-hardness vs.\ in 
$\PT^\Cmf$ are not tight bounds. The only exception is if $\Cmf=\PT$, where we again have that $\PT^\Cmf=\PT$.

To understand the complexity of finding justifications, several ontology languages with lower-complexity entailments---mainly
polynomial---have been studied. Unfortunately, the picture that arises from these studies is more complex than what is observed
in Proposition~\ref{prop:one:same}. 
For example, there are ontology languages, such as the language of propositional formulas in CNF, whose entailment problem
is \NP-complete, but \isjust is $D^P$-complete~\cite{PeSe17}.%
\footnote{$D^P$ is the class of problems which can be solved by one \NP and one co\NP test. It is believed to be strictly contained
in $\Delta^P_2=\PT^\NP$.}

Before considering the computation of all justifications, note that there are many important variants of \isjust which
may be considered. As we will see later, it is sometimes relevant to order the justifications according
to some preference (e.g., size) and \isjust becomes the problem of deciding whether a sub-ontology is the \emph{most preferred}
justification. This, of course, requires additional tests, and the complexity may change accordingly. For a study on how 
the complexity is affected by the preference relation and the ontology language see~\cite{PeSe17}.

\subsection{All Justifications}

If we want to solve \alljust, we can once again consider the black-box algorithm described in the previous section. For each
of the input sets $\Mmc_1,\ldots,\Mmc_n$, we verify in $\PT^\Cmf$ that they are justifications, and afterwards we need to verify
that no other set is a justification. The latter task can be performed through a \PSpace enumeration of all possible sub-ontologies,
checking for each of them that they are not a justification. Hence, overall, we can solve \alljust with a $\PSpace^\Cmf$ algorithm.
As in Proposition~\ref{prop:one:same}, the \PSpace base method can be absorbed into the oracle \Cmf if the latter is at least
\PSpace itself.

\begin{proposition}
\label{prop:all:same}
If entailment is \Cmf-complete for \Cmf at least \PSpace, then \alljust is \Cmf-complete as well.
\end{proposition}
While this proposition is very similar to the case for one justification, the \PSpace base at the algorithm makes the gap between
the lower and upper bounds, for ontology languages having simpler entailment tests, larger. Indeed, while for deciding whether
a set is a justification we incur in a jump at most of one level in the polynomial hierarchy, for \alljust the increment goes all the
way to the limit of this hierarchy at once. The gap can be reduced by considering the following idea: if an input is not an instance
of \alljust, we can verify it by guessing (in \NP) a new set \Nmc which does not contain any of the $\Mmc_i$s and verifying
(in \Cmf) that $\Nmc\models c$. Thus, \alljust is in co$\NP^\Cmf$.

The landscape of complexities also gets more complex in this case. There exist ontology languages with polynomial entailment
problems for which \alljust is polynomial, co\NP-complete, or hard for an intermediate class, respectively. There is also a
language with \NP-complete entailment problem for which the exact complexity of \alljust is unknown.

\begin{figure}[tb]
\centering
\includegraphics[width=0.9\textwidth]{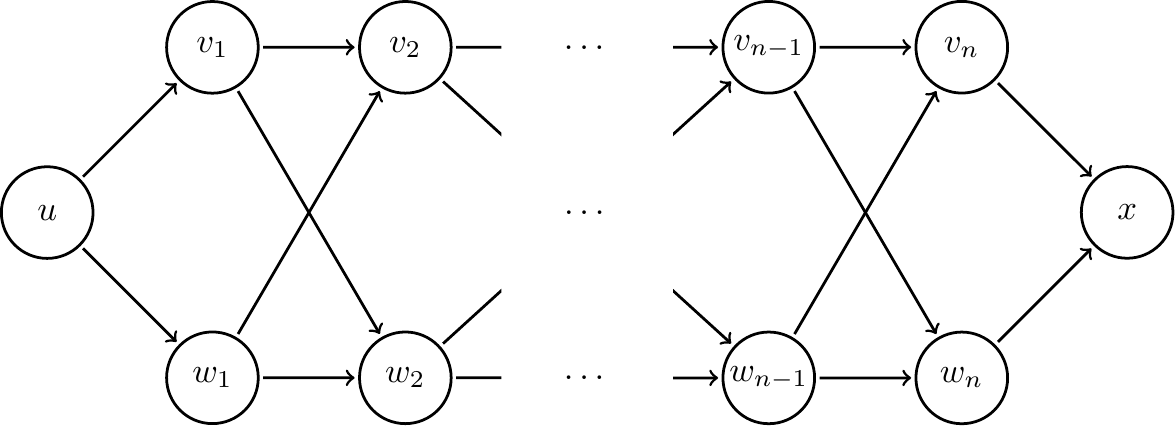}
\caption{A graph which produces exponentially many justifications.}
\label{fig:exp}
\end{figure}

So far, the discussion has focused on deciding \alljust, but in fact we are more interested in being able to enumerate all the justifications
(rather than deciding whether a set of sub-ontologies is indeed the class of all of them). 
The first thing to notice when dealing with the enumeration of justifications is that it is impossible spend less than exponential time
on this task. Indeed, even for the very simple ontology language that we are using as an example, it is easy to build an example
of a consequence that has exponentially many justifications w.r.t.\ an ontology~\cite{BaPS-KI07}. For example, there are
$2^n$ justifications for $(v,x)$ w.r.t.\ the graph from Figure~\ref{fig:exp}.
As before, this means that a full enumeration requires at least exponential time (but only polynomial space found justifications are
not preserved in memory). Thus, for expressive ontology languages, the black-box algorithm for computing all justifications 
is optimal in terms of complexity.

When dealing with enumeration problems, one can consider a different alternative complexity classes which take into account 
also the \emph{total} number of answers, and the time needed to obtain new answers~\cite{JoPY88}. Alternatively, one may try
to count the number of justifications available~\cite{Vali79}. Also in this case, the enumeration complexity varies with the specific 
ontology language, and whether a specific ordering is requested. For ontology languages related to directed graphs and hypergraphs,
the justifications can be enumerated with \emph{polynomial delay}; that is, allowing polynomial time (on the size of the ontology)
between two successive answers if no order is required, but may require total exponential time, but polynomial in the number of
justifications, if an order is required. For the light-weight description logic \EL, on the other hand, there is no algorithm that 
can enumerate all justification in total polynomial time. As a consequence, there must exist consequences with polynomially many
justifications, for which the enumeration problem requires super-polynomial time. For counting, in all the ontology languages 
with polynomial-time entailments studied so far, the complexity is \#P-complete while for propositional logic, where entailment is 
\NP-complete, counting the number of justifications becomes \#\NP-complete.

\section{Glass-box Algorithms}
\label{sec:gb}

We have seen that black-box algorithms are complexity-optimal for expressive ontology languages,
including the most common description logics---specifically, for any DL with value restrictions---assuming that an optimal implementation
of the standard reasoning method exists. In addition, these algorithms are
easy to implement and, as they only require repeated calls to an unmodified reasoner, can keep up to date with the newest 
optimisations and improvements. For that reason, they have become the approach of choice in those settings; for example, the
explanation service in Prot\'eg\'e~\cite{protege-owl} is implemented as a black-box. Nonetheless, it is important to study more 
targeted approaches, which have the potential of behaving better in practice. This is particularly true for ontology languages whose
complexity of reasoning is strictly below \PSpace.

The so-called \emph{glass-box} approaches to axiom pinpointing are based on a modification of the original reasoning algorithms
to be able to identify the justifications. In a nutshell, any reasoning algorithm must at some point consider the axioms in the 
ontology to decide whether the consequence follows. The idea of a glass-box algorithm is to trace which axioms were used
during this process, yielding candidates for justifications, or the whole class of justifications, depending on how the tracing
mechanism is implemented.

In DLs, the first proposals for a black-box algorithm for pinpointing were based on a modification of the tableaux-based 
reasoning method for reasoning in \ALC, which was originally designed for defeasible reasoning~\cite{BaHo95}. The idea
was slowly improved and generalised to allow for more expressive languages~\cite{ScCo-03,MLBP06,LMPB06} until a general 
approach for transforming tableaux-based reasoning methods into pinpointing algorithms was developed in~\cite{BaPe10b}.
The main drawback observed in these glass-box proposals is that, to guarantee that all the relevant axioms have been traced,
important optimisations have to be disabled. For example, for standard reasoning it suffices to halt the execution of the tableaux
when the desired consequence is obtained, but for pinpointing one must continue until \emph{all} possible derivations have been
explored. As a consequence, these pinpointing extensions do not behave well in practice. An additional disadvantage highlighted
in~\cite{BaPe10b} is that the pinpointing extension of a terminating tableau algorithm is not guaranteed to terminate in general.
On the other hand, some basic properties which are satisfied by all DL tableau algorithms suffice for guaranteeing termination.

Following a different approach, \cite{BaPe10} introduced a pinpointing approach based on weighted automata. Briefly, the 
approach takes an existing automata-based reasoning method, and transforms it into a pinpointing approach by adding weights,
belonging to the free distributive lattice, to all the transitions of this automaton. The \emph{behaviour} of the weighted automaton
obtained this way is a compact representation of all the justifications of the consequence. The main advantage of this approach
is that it is optimal in terms of worst-case complexity. However, as in standard automata-based reasoning, the \emph{best-case}
complexity also matches the worst case, making it impractical for real applications. 

To-date, the most successful approach which could be called glass-box is based on a translation of the execution of a 
consequence-based
algorithm into a propositional formula. In essence, one tries to reduce an axiom pinpointing problem to pinpointing in a well-known
ontology language for which efficient implementations exist. 
The original idea was introduced in~\cite{SeVe-cade09,SeVe-tr15} for the light-weight
DL \EL, but can be easily generalised to other consequence-based algorithms. Very briefly, consequence-based algorithms
work by applying rules over the explicitly represented knowledge (originally, the ontology) to make some of its implicit
consequences explicit. The method from~\cite{SeVe-cade09} introduces a new propositional variable for each consequence,
and a Horn clause simulating each possible rule application. In addition, it adds the representative variable of each axiom of the original
ontology.

As a very simple example, consider our graph ontology language, where the entailment relation is reachability. In this case, 
a consequence-based algorithm would only have the rule $\{(X,Y),(Y,Z)\} \to (X,Z)$ expressing that if we have the (explicit)
knowledge that the ontology entails $(X,Y)$ and $(Y,Z)$, then we can derive that it also entails $(X,Z)$. From this rule, and
the graph in Figure~\ref{fig:exagraph}, we obtain a set of Horn clauses which contains, among others,
\begin{align*}
x_{(u,v)}\land x_{(v,w)} \to x_{(u,w)}, &&
x_{(u,v)}\land x_{(v,x)} \to x_{(u,x)}, &&
x_{(u,v)}\land x_{(v,y)} \to x_{(u,y)}, \\
x_{(u,x)}\land x_{(x,w)} \to x_{(u,w)}, &&
x_{(u,x)}\land x_{(x,y)} \to x_{(u,y)}, &&
x_{(u,y)}\land x_{(y,u)} \to x_{(u,u)}.
\end{align*}
The ontology is then represented through the variables
\begin{align*}
x_{(u,v)}, &&
x_{(u,w)}, &&
x_{(v,w)}, &&
x_{(v,x)}, &&
x_{(v,y)}, &&
x_{(x,w)}, &&
x_{(x,y)}, &&
x_{(y,u)}.
\end{align*}
The conjunction of all these elements yields a Horn formula $F$ that entails the variable representing each relevant consequence.
Hence, for example $F\models x_{(u,w)}$. 

In order to find the justifications for a consequence $c$ w.r.t.\ the ontology \Omc, one needs only to enumerate the justifications
for the consequence variable $x_c$ w.r.t.\ the set of Horn clauses $F$, with the difference that the Horn clauses simulating the
execution of the rules are always present; that is, a justification may only remove from the formula variables representing the 
original ontology. 

In essence, what this translation does is to reduce axiom pinpointing in an arbitrary ontology language which accepts 
consequence-based reasoning into axiom pinpointing in propositional logic. The advantage is that we can then focus on the
development of highly-efficient pinpointing tools for this very specific ontology language, which can be further optimised taking
the shape of the obtained problem into account. Indeed, this idea has given rise to many axiom pinpointing tools for an extension
of \EL~\cite{AMIMPM16,MaPR16,SeVe-tr15} of which the most efficient---and only system capable of computing all the 
justifications for all the 5415670 atomic entailments from the very large ontology {\sc Snomed} is PULi~\cite{KaSk18}.

\subsection*{The Grey-Box Approach}

Before looking into different applications and variations of axiom pinpointing, we would be amiss to ignore a popular approach 
for finding \emph{one} justification, which combines the glass-box tracing approach with the black-box method from
Algorithm~\ref{alg:one}; hence, it is usually called \emph{grey-box}. The main idea, as with the glass-box approach, is to 
trace all the axioms used to prove the entailment of a consequence, through the application of an algorithm;
for example, in a tableau-based or consequence-based algorithm. However, in contrast to the case for finding all justifications,
we stop once the consequence is derived. This yields a set of axioms that is not guaranteed to be a justification, but from which
the consequence still follows. That is, it might still require some pruning of superfluous axioms to become a justification. On the
other hand, this limited tracing only imposes a minimal overhead to the original decision algorithm, and does not affect the existing
optimisations. Moreover, the resulting set tends to be much smaller than the original ontology, and very close to a justification.
Once this smaller ontology is found, it can be minimised by a call to the black-box method, yielding a justification. Note that this
second step needs to check potentially much less axioms than a direct use of Algorithm~\ref{alg:one} on the original ontology.

This grey-box approach was used originally by the CEL system~\cite{BaLS-IJCAR06} to find justifications efficiently for \EL, 
where the goal was not to find them all, but only a relevant subset. A similar idea is followed in the database community to trace 
the facts which provide an answer to a query. Although, to the best of our knowledge, there is no implementation of the grey-box 
approach for more complex ontology languages, or based on other kinds of algorithms, it should be relatively straightforward to 
modify a tableaux-based tool for this purpose.

An important drawback of the grey-box algorithm as described above is that the sub-ontology obtained by the tracing step is
only guaranteed to contain \emph{one} justification. This means that if one is interested in potentially finding more justifications,
then the whole process needs to be started anew for each successive solution. To alleviate this problem, some work has focused
on computing \emph{justification-preserving modules}; that is, sub-ontologies which are still guaranteed to contain all the 
justifications for a given consequence. Ideally, these modules should be fast to compute and still be small enough to guarantee that 
all justifications can be extracted from them following, e.g.\ the black-box approach, efficiently. It has been observed that
different modularization methods yield adequate solutions in this direction. Usually, these modules are based on syntactic
or semantic relationships between the axioms of the ontology \cite{StPS09,CHKS07,Sunt-ESWC08}. 
More recently, a modularization method based on a
modification of the reasoning algorithm was proposed \cite{PMIM17,KuLM06,MKIM19,MIMML14}. 
These approaches modify the tracing technique of the glass-box method. As usual, they keep track of the axioms that
are being used during the execution of the algorithm for deriving new knowledge, but instead of distinguishing between
different derivation paths, they are all stored in one single set. This simple modification avoids the problems with termination
of the original glass-box approach, but provides more information than just following one derivation as in the description
of the grey-box approach above. In empirical analyses, this approach has shown to be very helpful for solving axiom-pinpointing
related tasks.

\section{Applications and Variations}

The task of axiom pinpointing as defined in Section~\ref{sec:ap} has been extended to cover a large class of
variations, and can be used to solve different reasoning tasks.

The simplest generalisation which we can consider is to allow some parts of the ontology to be fixed; that is, consider a 
justification to be a minimal subontology \Mmc that, when added to a fixed ontology $\Omc_\mathsf{f}$, entails the consequence.
This application makes sense in the context of debugging, when we trust some axioms, and we do not want to look into them
when trying to understand or correct an erroneous consequence. We have also observed its use in the reduction from
consequence-based algorithms to axiom pinpointing in SAT in Section~\ref{sec:gb}. Further generalising this idea, one can
consider the atomic elements to be not axioms, but rather sets of axioms. In the literature, these have been called
group-MUSes \cite{LiSa08} or \emph{contexts} \cite{BaKP12}. Note that these contexts may appear implicitly in several
applications. For example, in consequence-based algorithms \cite{Sima:PhD,SiMH14}, a common pre-processing step is to modify 
the ontology into a given normal form. In those cases, several separated axioms may have been produced by a single original 
statement, and thus should all be considered together. Another setting where considering sets of axioms makes sense is
to find classes of modules of different kinds where a consequence can be derived. In this case, we can group axioms according
to the atomic decomposition of the ontology \cite{Ves+13}.

Returning justifications at the granularity of (sets of) axioms from the original ontology is not always satisfactory. These may
hide some relevant information or, conversely, be too complex to be understood or managed adequatedly by domain 
experts. Hence some research---specially in DLs---has considered building different representations. 
One direction produces so-called \emph{laconic} and \emph{precise} justifications \cite{HoPS-ISWC08} which, from 
a very abstract point of view, provide only the specific pieces of the axioms which are really responsible for the consequence, 
removing any superfluous information. The other direction combines several axioms into \emph{lemmas}, which remove 
excessive detail and become easier to read \cite{HoPS09}.

In a similar direction, more recent work has focused on the goal of minimally modifying ontologies through weakening
\cite{TGPPK18,BKNP18}. Very briefly, after a justification has been found, its axioms are modified to weaker ones in order
to get rid of the consequence. Hence, this process goes one step beyond pinpointing by further identifying the strongest
possible weakenings which yield the desired result. Note that axiom pinpointing is a special step of this idea, where the
only possible weakening an axiom is to replace it by a tautology.

Regarding the computation of all justifications, it is sometimes convenient to try to enumerate them in a specific order. 
For example, one may want to observe the smallest (w.r.t.\ cardinality) justifications first, or alternatively have a 
pre-specified order for accessing them. Obviously, once that we can compute all the justifications, it is also possible to 
order them before showing them to the user, but this brings a large overhead in general. To understand the issue better,
the complexity of enumerating justifications w.r.t.\ some natural orderings has also been studied, with the unsurprising
result that it depends not only on the ontology language, but also on the chosen order \cite{PeSe17}.

Axiom pinpointing is not only useful for understanding and potentially correcting consequences from an ontology, but 
it has also found applications aiding different kinds of supplemental and non-standard reasoning tasks. Perhaps the most 
studied and applied to date is related to the representation and handling of uncertain knowledge. When dealing with 
probabilities, several semantics and applications use axiom pinpointing, albeit often implicitly. In probabilistic logic programming
\cite{RaKT07}, the probability of an inference is given by the probabilities of combinations of facts that entail it (together 
with a fixed program) which correspond to the justifications of an ontology. This idea was generalised to description logics
under the so-called Disponte semantics \cite{RBLZ15}. In these approaches, it is assumed that all the uncertain elements are
probabilistically independent. To lift this assumption, newer approaches include a Bayesian network which expresses the 
joint probability distribution of the axioms (or more in general, contexts) of the ontology \cite{CePe17,BoMP19,dAFL08}.
For all these approaches, axiom pinpointing provides a helpful intermediate step which can be exploited also for approximating
solutions. For example, finding the justification with the highest probability can often yield a good approximation for the full
probability. In terms of possibility theory, when using the standard min-max semantics, it is well known that it suffices to
find one justification with the highest possibility degree \cite{BeBo17}.

Another application is in the context of access control, where one wants to provide only partial views to an ontology
to different users \cite{BKP09:latticetrustreasoning}. Here, axiom pinpointing is not only useful to find out the
access levels (that is, the contexts) where a consequence is derivable, but also to suggest changes in the access level
of axioms in order to hide or open implicit consequences to some users \cite{KnPe10}.

There is a current interest in being able to reason in the presence of inconsistencies or errors in
an ontology \cite{BeHS-IT05,LLRRS10,BiBo-RW16,LuPe14}. In this case, we are interested in finding so-called \emph{repairs},
which correspond to the dual notion of a justification; that is, maximal subontologies which do not entail the consequence
under consideration. Axiom pinpointing comes into place in this case in two different ways. On the one hand, repairs can
be computed from justifications through a hitting set computation. On the other hand, most of the techniques developed
for axiom pinpointing (and in particular, the black-box methods) can be easily adapted to compute repairs directly, usually
leading to the same complexity bounds \cite{Pena:PhD}. Still, keeping this dual view is often helpful for finding new
research problems.

To finish this brief overview on applications, we mention \emph{provenance} 
\cite{Green07-provenance-seminal,Geerts16-provenance,Dividino2009,CLOPX19}, which has a similar motivation,
but different development, to axiom pinpointing. Originally developed for databases, and later extended to DLs,
the goal of provenance is to track the origins for an answer to a query, in terms of the facts used to derive it. The main
difference with axiom pinpointing is that with provenance one is also interested in finding how often an axiom (or group of
axioms) is used in the derivation, and that minimality is replaced by a weaker notion, where all axioms need to be relevant,
but they may be superfluous.

\section{Conclusions}

We have presented a general overview on axiom pinpointing: what it is about, how it is usually tackled, and its main applications
and variations. We insist here that, although most of the terminology used was originally introduced in the context of 
description logics, we have tried to keep the presentation as general as possible to include the notions developed in many
other areas of knowledge representation and reasoning. Our hope is that this general presentation helps as a first step
towards a unified view of the problem of finding the axiomatic causes for consequences in different languages, and ultimately
leads towards an exchange of techniques, problems, and tools between areas. 

Clearly, the work on axiom pinpointing does not finish with this content. There exist many 
applications and variations to the problem which have not been mentioned, or commented only briefly, and which would 
require much more space to explore in detail. Even for the ones which were more thoroughly commented, there are 
usually many open problems which still require some additional work. Moreover, even at its basic state, axiom pinpointing
is not fully resolved. For example, in terms of its practical application, the development of efficient tools for more expressive
languages---or, even better, for general use---is still missing. 

There are many specialised techniques developed in areas other than description logics (e.g., databases, CSP, and 
SAT), which may be applicable in other areas with minor modifications. Alternatively, it may be possible to reduce problems
between areas as done in the reduction to propositional logic described here. This allows to focus on one specific problem,
and using advanced engineering techniques for its effective solution.

\bibliographystyle{splncs04}
\bibliography{bib}

\end{document}